# Data Analytics-enabled Intrusion Detection: Evaluations of ToN_IoT Linux Datasets

Nour Moustafa*, Mohiuddin Ahmed†, and Sherif Ahmed‡
School of Engineering and Information Technology, University of New South Wales, Canberra, Australia *
School of Science, Edith Cowan University, Perth, Australia †
School of Computer Science, University of Windsor, Windsor, Canada ‡
Email: *nour.moustafa@unsw.edu.au, † m.ahmed.au@ieee.org, ‡ sherif.saadahmed@uwindsor.ca

*Abstract*—With the widespread of Artificial Intelligence (AI)-enabled security applications, there is a need for collecting heterogeneous and scalable data sources for effectively evaluating the performances of security applications. This paper presents the description of new datasets, named ToN_IoT datasets that include distributed data sources collected from Telemetry datasets of Internet of Things (IoT) services, Operating systems datasets of Windows and Linux, and datasets of Network traffic. The paper aims to describe the new testbed architecture used to collect Linux datasets from audit traces of hard disk, memory and process. The architecture was designed in three distributed layers of edge, fog, and cloud. The edge layer comprises IoT and network systems, the fog layer includes virtual machines and gateways, and the cloud layer includes data analytics and visualization tools connected with the other two layers. The layers were programmatically controlled using Software-Defined Network (SDN) and Network-Function Virtualization (NFV) using the VMware NSX and vCloud NFV platform. The Linux ToN_IoT datasets would be used to train and validate various new federated and distributed AI-enabled security solutions such as intrusion detection, threat intelligence, privacy preservation and digital forensics. Various Data analytical and machine learning methods are employed to determine the fidelity of the datasets in terms of examining feature engineering, statistics of legitimate and security events, and reliability of security events. The datasets can be publicly accessed from [1].

## I. INTRODUCTION

Computer systems are without a doubt, an integral part of everyday life. Along with the prevalence of the Internet of Things (IoT) and Industrial IoT (IIoT) systems, the number of deployed devices is ever increasing. Ensuring computer systems' stability, protecting them against cyber threats, achieving their integrity, and maintaining their confidentiality is of utmost importance. As with many other computer systems, the IoT makes use of the Internet, to send data to a Cloud system, from which users can view diagnostics or any information recorded by their devices [2]. IoT devices are not properly secured, with their designers often neglecting to fortify them against attacks, due to added costs and the constrained resources of smart things [4], [5].

This work was funded by Australian Research Data Commons (*RG192500*) and UNSW Canberra (*PS51776*). Free use of the TON_IoT datasets for academic purposes is hereby granted in perpetuity. Use for commercial purposes is allowable after asking the author, Dr Nour Moustafa, who has asserted his right under the copyright.

Cyber-attacks have shown an interest in targeting IoT/IIoT systems, as they are easy to exploit, have some processing capabilities and are often in an "always-on" mode, providing the attackers with a consistent platform from which they can then launch further attacks. Computer devices are outfitted with an Operating System (OS), which manages its hardware and provides services to its programs. Although there exist multiple types of OSs, in this paper we focus on Linux OS. Linux is a Unix-like, open-source OS, which has been widely used in personal computers, servers and other commercial devices in the form of embedded OSs [6], [7]. With the emergence of the IoT, the popularity of light-weight Linux-based OSs has further increased. As such, investigating cyber threats for Linux OSs and devising methods for improving their security is an important task.

Because of the multiple threats that may target computers, many defensive mechanisms have been developed over the years [8], [9]. A group of such mechanism which has seen much development in recent years, are Intrusion Detection Systems (IDSs). Depending on their focus, IDSs are specialized into two groups: Host and Network IDS [8]. An Network IDS (NIDS) focuses on the network aspect of a system, monitoring inbound and outbound traffic from strategic locations, where it is installed. On the other hand, a Host IDS (HIDS) is installed on individual hosts in a network and monitor a device's internal state. Furthermore, depending on the methodology used to detect an attack, they are further categorized into signature-based detection, anomaly-based detection or hybrid of the two. Signature-based IDS maintains a database of known attacks and patterns that identify them, while anomaly-based IDS are trained to identify the normal behavior of the legitimate user, flagging any deviations as attacks [9]. To develop and evaluate IDSs, it is necessary to use high-quality data that represent realistic and current normal and attack events.

Existing datasets that can be used for the development of a HIDS, face several issues. To begin with, most of the existing Linux-derived datasets are not evaluated within an IoT environment [10], [11]. This is a serious drawback, as it is increasingly normal to find smart devices deployed in both personal and public networks, and their security weaknesses threaten the security of their local networks. Furthermore, most datasets intended to be used for HIDSs, focus solely on system calls [10], ignoring traces found in memory (RAM), the hard disk (HD) and the CPU which may cause some attacks to go unnoticed. Finally, some datasets lack credibility, either because they lack ground truth, or the analysis provided to describe them is poor.

This paper seeks to address the issues above, by introducing a new Linux-based dataset that does not focus solely on system calls, but further considers traces found in the memory, process and hard disk. The dataset was generated from a new testbed, which incorporated IoT smart thing traces in the normal data, and applied up-to-date attack techniques. The main contributions of this paper are as follows. First, a Linux-derived dataset is proposed that includes a wide range of activities related to hard disk, processes, memory, and network was generated, for the training and validation of IDS. Second, new features are employed that indicate links with real networks, such as Service Orchestration (SO), Software Defined Network (SDN) connectivity and Network Function Virtualization (NFV). Third, various machine learning models are utilized to evaluate the reliability of the Linux-derived datasets, along with the authentic ground truth also provided.

The structure of this paper is as follows. Section II explains the related work. Section III illustrates the testbed that was designed, while Section IV describes the Linux-based datasets. Section V presents the results of the statistical analysis and Machine Learning (ML) evaluations. Finally, in Section VI concluding remarks are given.

## II. BACKGROUND AND RELATED WORK

In this section, background information related to Host-based Intrusion Detection Systems (HIDSs), their variations, and use-cases are presented. This is followed by several existing datasets that have been developed to construct and evaluate IDSs.

### A. An overview of HIDS

Depending on which part of the network an IDS is applied, and what types of data it processes, there exist two main categories of IDS: network-based and host-based IDS [8], [12]. Network IDS are placed at strategic points of the network, monitoring inbound and outbound traffic to identify suspicious network traffic. On the other hand, host-based IDS are installed in individual hosts, monitoring their internal mechanism to detect the presence of unauthorized actions. HIDS can be separated into three main subcategories, with respect to the detection techniques they employ, namely signature-based, anomaly-based detection and hybrid of the two types. Signature-based HIDS [13] relies on a knowledge database comprised of character sequences that identify attacks, called signatures. The concept of using signatures for attack detection was originally applied to antivirus software, where files were scanned for known patterns of malware.

Signature-based HIDS scan logs, memory dumps and network traffic originated or received by the host where it has been installed. In general, they are characterized by having a high detection rate for known attacks while maintaining a high detection speed, although they face issues if the slight alterations are introduced to the attack sequence/code. Furthermore, they are unable to detect zero-day exploits [14]. On the other hand, anomaly-based HIDS [14] works by establishing a profile that describes normal actions performed by a legitimate user, identifying any deviations from that profile as an intrusion. They employ machine learning and deep learning mechanisms to learn patterns of legitimate behavior, thus they can detect unknown (zero-day) attacks something which can not be accomplished with signature-based HIDS. Furthermore, as anomaly-based HIDS does not rely on signatures, they can detect attacks that have been slightly altered (called a mutation), a process that hackers use to overcome signature-based HIDS. However, because they employ machine or deep learning models, they are more computationally intensive than signature-based [14], [15]. Besides, anomaly-based detection techniques would produce higher false alarm rates, when detecting known patterns, compared to the signature-based detection techniques, due to the generalization that is required for machine learning techniques to make predictions about unknown data.

### B. Datasets used for evaluating HIDS

Since the design, generation and evaluation of an IDS relies on the data used, generating a reliable and up-to-date dataset is of utmost importance. Over the years, several datasets have been developed to develop cyber-security tools. Although significant research has focused on network-related incidents [16], datasets used for HIDS have also been developed. The most commonly used datasets are briefly discussed below.

- *The DARPA 98 dataset* [17] is considered to be the first attempt to generate a dataset for the creation of IDS. The DARPA 98 dataset was generated in seven weeks by the MIT Lincoln Laboratory. It is comprised of raw pcap files at a size of 4GB. This dataset is outdated [18], [19] as attacks have become more intricate since 1998 and newer technologies are in use today, like the IoT. Furthermore, the dataset focused on network incidents, thus ignoring host-related data that could be used by a HIDS.
- *The SSENet-2014 dataset* [20] was derived from the SSENet-2011 dataset. The newer dataset is comprised of 28 attributes with 9 basic, 9 network traffic and 10 host attributes. Contrary to the other datasets, SSENet-2014 was generated by attacking a vulnerable Windows server, instead of a Linux machine. The derived attributes match those of the KDD-99 dataset. Not much information is given about the statistics of the dataset.
- *The ADFA-LD dataset* [10] is a Linux-derived dataset intended for anomaly-based HIDS generation and validation. The testbed included an Ubuntu OS machine, outfitted with various network-enabled programs, which was the target of several attacks, including password bruteforce, privilege escalation and Meterpeter-generated payloads. During collection, the Ubuntu OS was scanned, with the final dataset comprised of three groups of raw system call data, one group for training on normal data, the second for validation on normal data and the last group comprised of attack data.
- *The NGIDS-DS dataset* [21] was generated by using the cyber-range at ADFA, Canberra. The dataset combines network traces, derived from the IXIA Perfect Storm tool and host-based data that were recorded from the cyber-range set-up. The testbed comprised of two Ubuntu OS machines, one used to collect network traffic and the other for collecting process-related data with a focus on system calls. The environment was engaged for approximately four days, resulting in more than ninety million records of both network traffic and host log information.

Although significant research has been conducted for the generation of datasets for IDS development, their drawbacks justify the need for the generation of a new Linux-based dataset. More importantly, some datasets include outdated attack and normal traffic scenarios, while others either focus entirely on process system calls, ignoring other sources of traces like the memory, hard drive, or the processor state. Finally, these datasets did not incorporate any IoT elements in their testbeds and thus lack IoT-derived traces. This paper seeks to address these shortcomings by proposing new Linux-based datasets in an IoT/IIoT environment.

## III. Suggested Testbed architecture for creating TON_IoT Linux dataset

The suggested testbed architecture of the ToN_IoT datasets for gathering the dataset of Linux systems is depicted in Figure 1. The architecture was designed using the network communication of Linux systems and IoT in the three layers of edge, fog and cloud to emulate the realistic execution of modern real-world IoT networks. The dynamics of the three layers, including physical and simulated environments, were flexibly controlled using the platforms technologies of SO, SDN and NVF. The NSX-VMware platform [22] was employed to offer an SDN and SO solution for the suggested testbed. This technology allows the generation of overlay networks with the same abilities as physical networks.

The VMware NSX platform was implemented to synchronously manage Linux operating systems and IoT services. In VMware NSX, the vCloud NFV platform was used to offer a modular design with abstractions that allow multi-domain, hybrid physical, and VM deployments [23]. The platform facilitates the design of a dynamic testbed network via managing various Virtual Machines (VMs) for running legitimate and malicious scenarios. This also allows the connectivity between the edge, fog and cloud services, as explained below.

- **Edge layer** - includes the physical devices and the operating systems employed as the infrastructure of constructing the virtualization technology and cloud services at fog and cloud layers. It involves several IoT devices, such as temperature sensors, smartphones and smart TVs, as well as host systems, including clients and servers (i.e., Linux OSs), utilized to link the IoT and network devices and systems to the Internet. The NSX-VMware

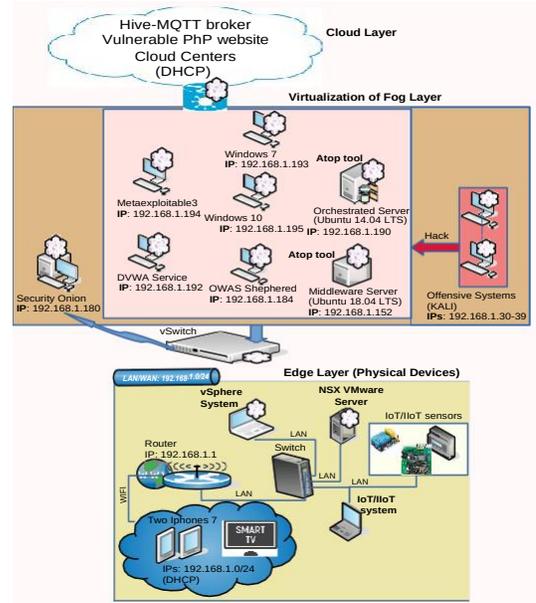

Fig. 1. Testbed Architecture of TON_IoT datasets for gathering Linux dataset

platform was configured on a host server to operate the VMs deployed at the fog layer.

- **Fog layer** involves the virtualization technology, which manages the VMs and their services using the NSX-VMware and vCloud platform. This platform implements the SDN and NFV services in the Linux testbed. It allows the dynamic testbed network of the Linux ToN_IoT via designing and managing various VMs for developing suspicious and legitimate scenarios. This layer involves the nodes of VMs configured to generate the datasets such as the orchestrated and middleware servers. The Linux data were collected from network, memory, process and hard drive of these servers.

- **Cloud layer** includes the cloud services developed online in the architecture. The fog and edge services were linked to the cloud HIVE MQTT broker [24], the public PHP vulnerable website [25], cloud virtualization, and cloud data analysis services such as services Microsoft Azure. The HIV MQTT broker helps in publishing and subscribing to the telemetry data of IoT services via the node-red tool. The PHP vulnerable website was employed to launch attacking activity against websites. The other cloud services, such as Microsoft Azure, were used to transfer telemetry data to the cloud and present their behaviors.

The main VMs configured at the fog layer for collecting Linux data are described in the following:

- *Orchestrated server*–is a VM server designed in the testbed using the Ubuntu 14.04 LTS with the IP address (*192.168.1.190*). The server facilitated various orchestrated services, for example, Kerberos, HTTPS, and DNS to simulate realistic network services and create simulated network traffic using the Ostinato Traffic Generator [26] that transfers network traffic to other systems in the testbed.
- *Middleware server*–is the IoT virtualized server implemented in the testbed using the Ubuntu 18.04 with the IP address (*192.168.1.152*). The server involved the scripts that execute IoT services through public and local MQTT brokers employed in the testbed and connected to the cloud layer for publishing and subscribing to the sensing data of IoT services.
- *Client Systems*–involve a Windows 7 VM (IP address: *192.168.1.193*), Windows 10 VM (*192.168.1.195*), DVWA web service (*192.168.1.192*), OWASP security Sphered VM (192.168.1.184), Metaspoitable 3 (*192.168.1.194*). The Windows VMs were employed to execute the web interface of the node-red IP (*192.168.1.152*). The Damn Vulnerable Web App (DVWA) [27] was used to make network weaknesses through web applications attacked by the offensive systems. The OWASP security Sphered VM [28] is an open-source platform that includes various weaknesses against mobile and web applications hacked by the offensive systems. As well, the Metasploitable3 VM [29] was configured to increase vulnerable fog nodes and exploit them by multiple hacking techniques.
- *Offensive systems*–contain the Kali Linux VMs and scripts of attacking techniques breach vulnerable systems. Ten IP addresses (i.e.,*192.168.1.30-39*) were used in the testbed to launch attacking techniques and breach vulnerable systems either IoT services (i.e., MQTT brokers and node-red), operating systems (i.e., Windows 7 and 10, and Ubuntu 14.04 LTS and 18.04 LTS), and network systems (i.e., IP addresses and open protocols of the VMs).
- *Data Logger Systems*–log data of the Linux operating systems included in the testbed (i.e., orchestrated and middleware servers). The atop tool [30] was used to capture the raw data of memory, process and hard disk of Linux OSs. This toot captures the most critical hardware resources on system-level, including cpu, memory, and disk. The raw data were then logged in a CSV format and training-testing CSVs for executing the AI-enabled security applications such as intrusion detection. The data features created for the Linux TONIoT dataset are explained below.

## IV. NORMAL/ATTACKING SCENARIOS AND DATA FEATURES

This section explains normal and attack scenarios included for labeling the dataset, as well as the data features extracted from memory, process, and hard disk of Linux OSs.

### A. Normal and Attacking Scenarios

Normal scenarios were employed various normal observations for the Linux dataset via configuring legitimate operating of the orchestrated and middleware servers with the clients without launching any hacking activity. For instance, publishing IoT sensing data between edge, fog and cloud layers, and sending network packets using the ostinato traffic generator between the VMs. These scenarios assisted in collecting large-scale Linux data from the Linux servers from their CPU, memory, and process.

Hacking scenarios were used to launch nine attack families against vulnerable systems of IoT services and operating systems. The scripts and some links of the attacking categories have been published in [1]. The attack families utilized in the datasets are explained below.

- *Scanning attack* - we utilized tools, such as the Nessus and Nmap, on the offensive systems with IP addresses (*192.168.1.20-38*) against the victim's subnet (*192.168.1.0/24*) and all other public vulnerable systems, such as nmap (*192.168.1.40-254*) for scanning open services in this IP range.
- *Ransomware attack* - we configured ransomware activity on the Kali Linux with IP addresses (*192.168.1.33,37*) to implement this malware against the weaknesses of systems involved in the testbed. The Metasploit toolkit was used to execute this attack such as exploiting the SMB vulnerability of the systems.
- *Denial of Service (DoS) attack* - we used various DoS attack techniques on the offensive systems with IP addresses (*192.168.1.30,31,39*) to breach vulnerable Linux services. We also developed Python scripts using the Scapy package to execute the variants of DoS attacks such as land and syn DoS ones.
- *Distributed Denial of Service (DDoS) attack* - we configured DDoS attacks on the offensive systems with IP addresses (*192.168.1.30,31,34,35,36,37,38*) to exploit many several weaknesses in the testbed and its Linux systems. Moreover, we developed bash scripts to launch execute several DDoS techniques against Linux and IoT weaknesses by using the ufonet program.
- *Injection attack* - we developed multiple injection attacks at the offensive systems with IP addresses (*192.168.1.30, 31, 33, 35*). We injected data inputs against web applications of DVWA and Linux VMs, for example, executing SQL injection, broken authentication, and unintended data leakage.
- *Backdoor attack* - we utilized the offensive systems with IP addresses (*192.168.1.33,37*) to assert the persistence of attacking using the Metasploit tool. This was developed by a bash script such as using the command "run persistence -h".
- *Password attack* - we employed the offensive systems with IP addresses (*192.168.1.30, 31, 32, 35, 38*) to launch password/bruteforce attacks. The hydra and cewl programs were used to execute password attacking events against vulnerable systems in the testbed.
- *Man-In-The-Middle (MITM) attack* - we used the offensive systems with IP addresses (*192.168.1.31,34*) to execute many MITM scenarios in the testbed. We configured the Ettercap tool to launch the scenarios, such as ARP spoofing, and port stealing.

- *Cross-site Scripting (XSS) attack* - we used the offensive systems with IP addresses (*192.168.1.32,35,36,39*) to illegally inject web applications of Linux, DVWA and Security Shepherd VMs. We developed suspicious bash scripts of Python codes to attack the web applications and Linux VMs of the testbed using the Cross-Site Scripter tool (named XSSer).

## B. Data features of ToN_IoT Linux dataset

The proposed Linux dataset was generated the Linux Oss (i.e., orchestrated and middleware servers), and incorporates collections of data from multiple sources, including the data of memory, process, and hard disk. The data was initially logged in a text format and then converted to a CSV format to ease its usage. To record system-related data, the atop program [30] was launched on the Ubuntu machines, which allowed the monitoring of the machines' various subsystems (process, memory, Harddisk). The final features of the Linux-derived dataset are presented in Tables I, II, and III. The features are each presented, along with their data type and a short description.

TABLE I
LINUX-DISK DATA FEATURES

| ID | Feature | Type | Description |
|---|---|---|---|
| 1 | PID | Number | Process identifier which is active in a Linux kernel |
| 2 | RDDSK | Number | Amount of data read from disk |
| 3 | WRDSK | Number | Amount of data written to disk |
| 4 | WCANCL | Number | Amount of data that was written but has been withdrawn |
| 5 | DSK | Number | Disk occupation percentage |
| 6 | CMD | String | process name which is active in a Linux kernel |
| 7 | label | Number | Tag normal and attack records, where 0 indicates normal and 1 indicates attacks |
| 8 | type | String | Tag attack categories, such as normal, DoS, DDoS and backdoor attacks, and normal records |

TABLE II
LINUX-PROCESS DATA FEATURES

| ID | Feature | Type | Description |
|---|---|---|---|
| 1 | PID | Number | Process identifier which is active in a Linux kernel |
| 2 | TRUN | Number | Number of threads in state running (R) |
| 3 | TSLPI | Number | Number of threads in state interruptible sleeping (S) |
| 4 | TSLPU | Number | Number of threads in state uninterruptible sleeping (D) |
| 5 | POLI | String | Scheduling policy (normal timesharing, realtime round-robin, realtime fifo) |
| 6 | NICE | Number | Nice value which is the more or less static priority that can be given to a proces on a scale from -20 (high priority) to +19 (low priority) |
| 7 | PRI | Number | Priority which is the process priority ranges from 0 (highest priority) to 139 (lowest priority). |
| 8 | RTPR | Number | Realtime priority which is according the POSIX standard. Value can be 0 for a timesharing process (policy norm, btch or idle) or ranges from 1 (lowest) till 99 (highest) for a realtime process (policy rr or fifo) |
| 9 | CPUNR | Number | Current processor which is the identification of the CPU the main thread of the process is running on or has recently been running on |
| 10 | Status | Number | Status of a process, where the first position indicates if the process has been started during the last interval (the value N means new process) |
| 11 | EXC | Number | Exit code of a terminated process (second position of column ST is E) or the fatal signal number (second position of column ST is S or C) |
| 12 | State | String | Current state of the main thread of the process |
| 13 | CPU | Number | CPU time consumption of this process in system mode (kernel mode), usually due to system call handling |
| 14 | CMD | String | The name of the process. This name can be surrounded by less/greater than signs (name) which means that the process has finished during the last interval |
| 15 | label | Number | Tag normal and attack records, where 0 indicates normal and 1 indicates attacks |
| 16 | type | String | Tag attack categories, such as normal, DoS, DDoS and backdoor attacks, and normal records |

As can be seen in the three tables of the data features, the last two features, label and type identify whether the record belongs to an attack or a normal instance, as-well-as its subtype. Furthermore, information between the three collections of data can be combined, by utilizing the pid feature, which uniquely identifies a process in a Linux system.

TABLE III
LINUX-MEMORY DATA FEATURES

| ID | Feature | Type | Description |
|---|---|---|---|
| 1 | PID | Number | Process identifier which is active in a Linux kernel |
| 2 | MINFLT | Number | The number of page faults issued by this process that have been solved by reclaiming the requested memory page from the free list of pages |
| 3 | MAJFLT | Number | The number of page faults issued by this process that have been solved by creating/loading the requested memory page |
| 4 | VSTEXT | Number | The virtual memory size used by the shared text of this process |
| 5 | VSIZE | Number | The total virtual memory usage consumed by this process |
| 6 | RSIZE | Number | The total resident memory usage consumed by this process |
| 7 | VGROW | Number | The amount of virtual memory that the process has grown during the last interval |
| 8 | RGROW | Number | The amount of resident memory that the process has grown during the last interval |
| 9 | MEM | Number | Memory occupation percentage |
| 10 | CMD | String | Process name |
| 11 | label | Number | Tag normal and attack records, where 0 indicates normal and 1 indicates attacks |
| 12 | type | String | Tag attack categories, such as normal, DoS, DDoS and backdoor attacks, and normal records |

Attack types include DoS, DDoS, Injection, MITM, Scanning, Password and XSS attacks, as discussed in Section IV-A.

## V. EXPERIMENTAL ANALYSIS

In this section, we discuss data analytics of features and results different machine learning algorithms to the proposed dataset. It is expected that the application of the machine learning algorithms will generate new insights to apply the TON_IoT Linux datasets for different AI-enabled security applications such as intrusion detection, privacy preservation and threat models.

### A. Data anaytics and Statistics of Linux dataset

The statistics of the three data collections that make up the Linux-derived dataset are presented in Figure 2. Specifically, the number of records of the disk, memory and process for the entire dataset (the top three tables) and the test-test (on the botton three tables) are displayed in the figure. The three data sets of disk, memory and process can be used in applying federated and distributed machine learning models as the three data sets are linked with the PID data features as shown in Tables I, III and II. They also have most of the attack events that have some vulnerabilities for exploiting hard disk, process and memory of the Linux systems.

### B. Machine Learning for evaluating Linux dataset

TABLE IV
PARAMETERS USED BY THE ANOMALY DETECTION ALGORITHMS

| Algorithm | Parameter |
|---|---|
| k-NN | $k$- number of neighbours considered. |
| LOF | $k_{min}$ and $k_{max}$: minimum and maximum number of neighbours. |
| COF | $k$- number of neighbours considered. |
| aLOCI | tree depth, number of grids, minimum neighbours and difference of levels. |
| INFLO | $k$- number of neighbours considered |
| LoOP | $k$- number of neighbours considered and normalization factor. |
| CBLOF | $\alpha$- percentage of normal data, $\beta$- ration between the sizes of clusters. |
| LDCOF | $\gamma$ - ratio between sizes of clusters. |
| CMGOS | $\gamma$, probability of normal class and covariance estimation. |
| HBOS | number of bins and binwidth. |
| RPCA | probability of normal class. |

In this section, we apply different machine learning algorithms to the proposed datasets using the *Rapid Miner* tool [44] on **the Linux_disk data only**. It is expected that the application of the machine learning algorithms will generate new insights. The experimental results will be able to identify

### linux_disk

| No of rows | Type |
|---|---|
| 71603 | ddos |
| 70688 | dos |
| 41321 | injection |
| 112 | mitm |
| 1610724 | normal |
| 51409 | password |
| 63745 | scanning |
| 17759 | xss |

### Linux_memory

| No of rows | Type |
|---|---|
| 45689 | ddos |
| 75283 | dos |
| 58523 | injection |
| 112 | mitm |
| 1891527 | normal |
| 26016 | password |

### Linux_process

| No of rows | Type |
|---|---|
| 71603 | ddos |
| 70721 | dos |
| 41311 | injection |
| 112 | mitm |
| 1636604 | normal |
| 51409 | password |
| 38449 | scanning |
| 17759 | xss |

### Train_test Linux_disk

| No of rows | Type |
|---|---|
| 10000 | ddos |
| 10000 | dos |
| 10000 | injection |
| 100000 | normal |
| 10000 | xss |
| 10000 | password |
| 10000 | scanning |
| 112 | mitm |

### Train_test Linux_memory

| No of rows | Type |
|---|---|
| 10000 | ddos |
| 10000 | dos |
| 10000 | injection |
| 100000 | normal |
| 10000 | password |
| 112 | mitm |

### Train_test Linux_Process

| No of rows | Type |
|---|---|
| 10000 | ddos |
| 10000 | dos |
| 10000 | injection |
| 100000 | normal |
| 10000 | xss |
| 10000 | password |
| 10000 | scanning |
| 112 | mitm |

Fig. 2. Statistics of data records of the Linux dataset

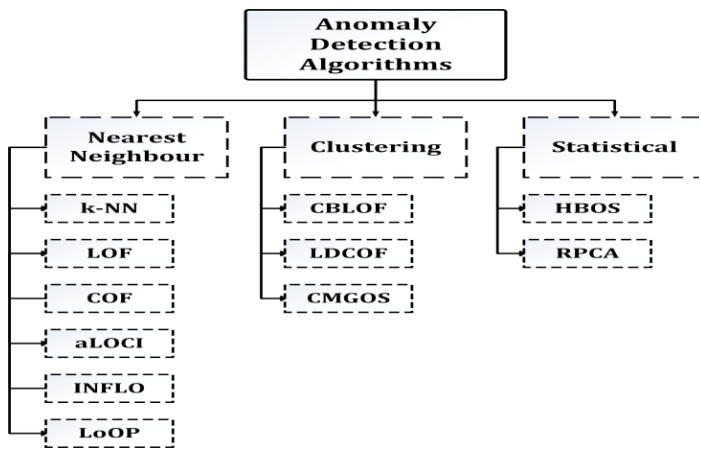

Fig. 3. Anomaly Detection Algorithms

TABLE V
PERFORMANCE OF ANOMALY DETECTION (HIT RATE) ALGORITHMS

| Algorithm | DDoS | DoS | Injection | Scanning | Password | Xss | MITM |
|---|---|---|---|---|---|---|---|
| k-NN | 100 | 53.5 | 61.03 | 84.94 | 23.56 | 89.44 | 99.1 |
| LOF | 38.6 | 50.39 | 53.47 | 50.72 | 33.18 | 57.64 | 74.1 |
| COF | 8.10 | 68.19 | 80.86 | 63.82 | 32.57 | 70.56 | 94.6 |
| aLOCI | 9.58 | 42.43 | 39.71 | 34.83 | 40.38 | 41.99 | 29.5 |
| INFLO | 8.84 | 42.75 | 50.04 | 25.33 | 38.57 | 47.17 | 58.9 |
| LoOP | 16.41 | 42.02 | 44.04 | 43.46 | 33.4 | 58.51 | 51.8 |
| CBLOF | 100 | 32.09 | 21.07 | 44.78 | 35.71 | 18.15 | 50 |
| LDCOF | 100 | 32.10 | 21.09 | 44.81 | 35.74 | 18.18 | 50 |
| CMGOS | 100 | 50.94 | 21.17 | 46.54 | 33.52 | 22.19 | 50 |
| HBOS | 100 | 64.57 | 51.63 | 66.22 | 23.57 | 78.77 | 66.07 |
| RPCA | 100 | 46.35 | 23.73 | 47.83 | 35.73 | 25.35 | 48.2 |

the effective algorithms for different types of attacks. This dataset contains seven different types of cyber attacks and we have used a standard set of machine learning algorithms to identify performance of these algorithms in detecting the different attacks. Figure 3 shows that the set of anomaly detection algorithms used are originated from three popular methods [41]–[43]. Given below are the anomaly detection techniques used for the experimentation:

- *k-NN*: A score for being anomalous is assigned to all the data instances based on the average distance to the nearest neighbours.
- *LOF*: All the data instances are assigned with an anomaly score based on the local density.
- *COF*: A variant of *LOF* based on density.
- *aLOCI*: Local correlation integral is used to assign the score for being anomalous to all data instances.
- *LoOP*: A probability score for being anomalous is given based on local density.
- *INFLO*: The concept of Influenced Outlierness based on neighbours are used to assign scores.
- *CBLOF*: The clustered data instances are assigned anomalous scores based on distances between larger and smaller clusters.
- *CMGOS*: The clustered instances are assigned scores based on their distances to the cluster center.
- *LDCOF*: Based on the distance to the nearest large cluster, scores are assigned for being anomalous..
- *RPCA*: Originated from principal component analysis.
- *HBOS*: A histogram based techniques that uses either fixed or dynamic binwidth to assign scores to all data

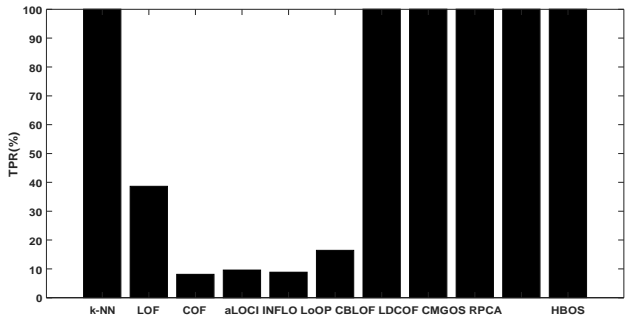

(DDoS)

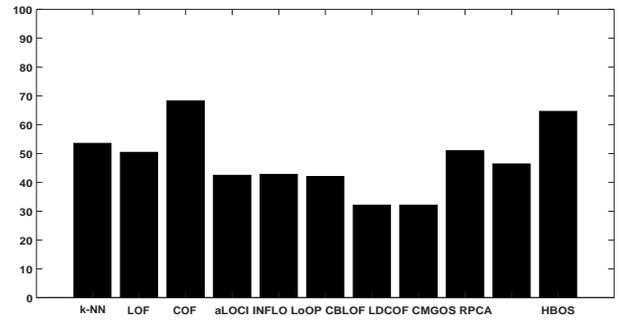

(DoS)

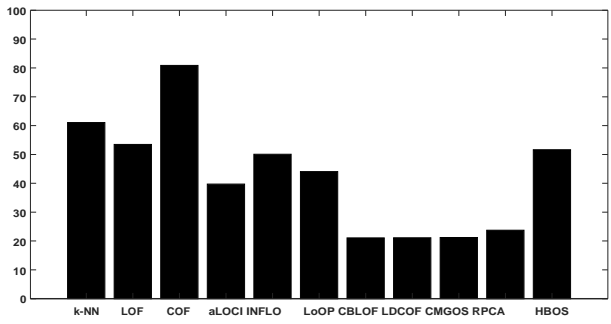

(Injection)

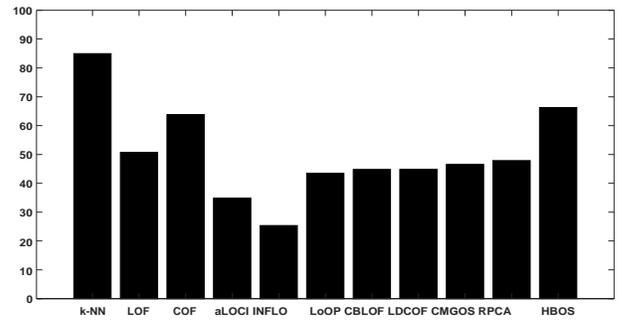

(Scanning)

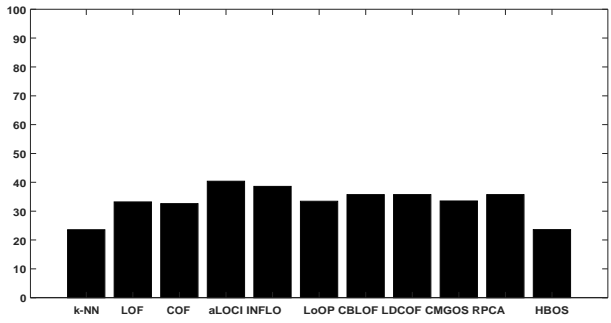

(Password)

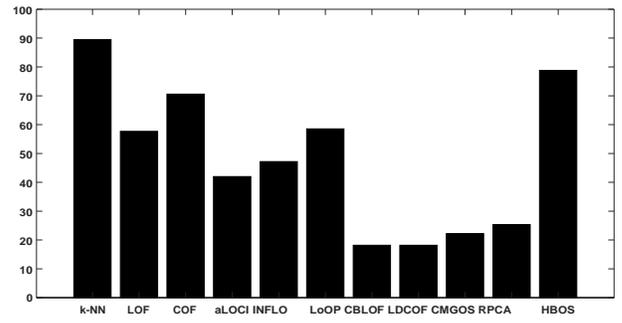

(XSS)

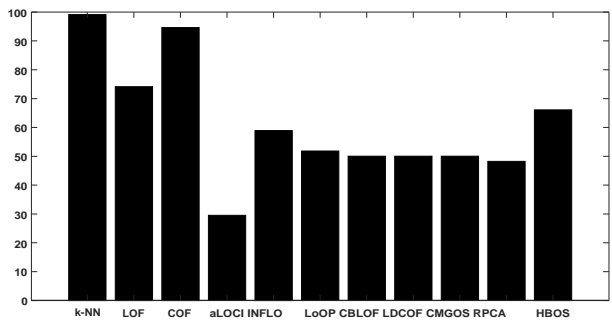

(Man in the Middle)

Fig. 4. Performance of anomaly detection algorithms on Linux dataset

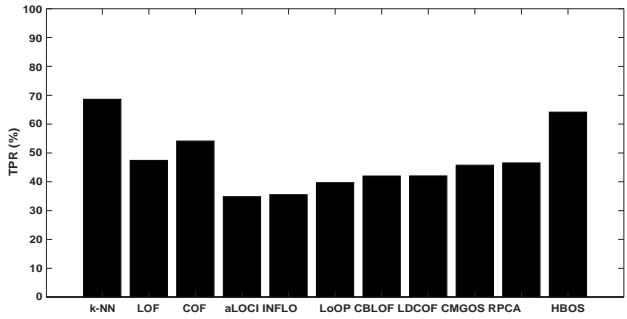

(Overall *TPR*)

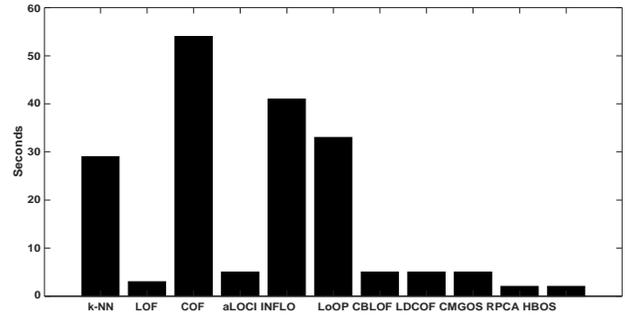

(Timing)

Fig. 5. Overall performance comparison

instances for being anomalous.

Table IV contains the parameters used the anomaly detection algorithms stated above. Table V[1] reflects the performance of the anomaly detection algorithms when applied to the dataset (Linux_disk). It is observed that the majority of the anomaly detection algorithms are able to identify the *DDoS* attacks with perfect hit rates (*True Positive Rate*), simultaneously some the techniques performed poorly on similar attack detection. For better understanding the results are also shown in Figure 4. In Figure 5, the overall hit rates of the algorithms and time required are shown. It is evident that *k-NN* and *HBOS* perform superior than others in terms of hit rates. However, *HBOS* is significantly better in terms of computational effectiveness (timing) than the classical *k-NN*. None of the algorithms had more than 70% hit rate and majority of them had less than 50%. Therefore, it is imperative that newer approaches are required to identify the attacks originated from state-of-the-art environment.

## VI. CONCLUSION AND FUTURE WORK

This paper has presented the description of the Linux TON_IoT datasets created at UNSW Canberra. A new testbed architecture was designed for collecting Linux data from memory, process and hard disk of Linux systems. The architecture involves a broad range of IoT services configured at the edge layer, virtual machines of operating systems implemented at the fog layer, and cloud services deployed at the cloud layer. The network connectivity between the layers was designed using the VMware NSX and vCloud NFV platform to offer SDN and NVF services. Modern legitimate and nine attack families were executed, with authentic ground truth that describes attack events for assessing the credibility of new AI-based cybersecurity systems. Several machine and deep learning models have been trained and validated using the Linux dataset. The results reveal that the dataset can be used to validate new AI-based cybersecurity applications, such as intrusion detection, malware detection, privacy preservation,

---

[1]Green cells represent the highest hit rates by an algorithm and red cells are for worst.

digital forensics, and threat intelligence, which we will investigate in the future.

## REFERENCES

[1] "Ton_ iot dataset," https://www.unsw.adfa.edu.au/unsw-canberra-cyber/cybersecurity/ADFA-ton-iot-Datasets/, July 2020.
[2] N. Koroniotis, N. Moustafa, and E. Sitnikova, "Forensics and deep learning mechanisms for botnets in internet of things: A survey of challenges and solutions," *IEEE Access*, vol. 7, pp. 61 764–61 785, 2019.
[3] G. Ho, D. Leung, P. Mishra, A. Hosseini, D. Song, and D. Wagner, "Smart locks: Lessons for securing commodity internet of things devices," in *Proceedings of the 11th ACM on Asia conference on computer and communications security*. ACM, 2016, pp. 461–472.
[4] E. Ronen, A. Shamir, A.-O. Weingarten, and C. O'Flynn, "Iot goes nuclear: Creating a zigbee chain reaction," in *2017 IEEE Symposium on Security and Privacy (SP)*. IEEE, 2017, pp. 195–212.
[5] Y. Seralathan, T. T. Oh, S. Jadhav, J. Myers, J. P. Jeong, Y. H. Kim, and J. N. Kim, "Iot security vulnerability: A case study of a web camera," in *2018 20th International Conference on Advanced Communication Technology (ICACT)*. IEEE, 2018, pp. 172–177.
[6] F. Edition, "Linux," 2016.
[7] A. Silberschatz, G. Gagne, and P. B. Galvin, *Operating system concepts*. Wiley, 2018.
[8] H.-J. Liao, C.-H. R. Lin, Y.-C. Lin, and K.-Y. Tung, "Intrusion detection system: A comprehensive review," *Journal of Network and Computer Applications*, vol. 36, no. 1, pp. 16–24, 2013.
[9] N. Moustafa, G. Creech, and J. Slay, "Big data analytics for intrusion detection system: Statistical decision-making using finite dirichlet mixture models," in *Data analytics and decision support for cybersecurity*. Springer, 2017, pp. 127–156.
[10] G. Creech and J. Hu, "Generation of a new ids test dataset: Time to retire the kdd collection," in *2013 IEEE Wireless Communications and Networking Conference (WCNC)*. IEEE, 2013, pp. 4487–4492.
[11] I. Sharafaldin, A. H. Lashkari, and A. A. Ghorbani, "Toward generating a new intrusion detection dataset and intrusion traffic characterization." in *ICISSP*, 2018, pp. 108–116.
[12] H. Kozushko, "Intrusion detection: Host-based and network-based intrusion detection systems," *Independent study*, 2003.
[13] S. Freeman, A. Bivens, J. Branch, and B. Szymanski, "Host-based intrusion detection using user signatures," in *Proceedings of the Research Conference. RPI, Troy, NY*, 2002, pp. 2005–2014.
[14] S. Jose, D. Malathi, B. Reddy, and D. Jayaseeli, "A survey on anomaly based host intrusion detection system," in *Journal of Physics: Conference Series*, vol. 1000, no. 1. IOP Publishing, 2018, p. 012049.
[15] M. Bijone, "A survey on secure network: intrusion detection & prevention approaches," *American Journal of Information Systems*, vol. 4, no. 3, pp. 69–88, 2016.
[16] N. Koroniotis, N. Moustafa, E. Sitnikova, and B. Turnbull, "Towards the development of realistic botnet dataset in the internet of things for network forensic analytics: Bot-iot dataset," *Future Generation Computer Systems*, vol. 100, pp. 779–796, 2019.

[17] R. P. Lippmann, D. J. Fried, I. Graf, J. W. Haines, K. R. Kendall, D. Mc-Clung, D. Weber, S. E. Webster, D. Wyschogrod, R. K. Cunningham *et al.*, "Evaluating intrusion detection systems: The 1998 darpa off-line intrusion detection evaluation," in *Proceedings DARPA Information Survivability Conference and Exposition. DISCEX'00*, vol. 2. IEEE, 2000, pp. 12–26.

[18] S. T. Brugger and J. Chow, "An assessment of the darpa ids evaluation dataset using snort," *UCDAVIS department of Computer Science*, vol. 1, no. 2007, p. 22, 2007.

[19] S. Zanero, "Flaws and frauds in the evaluation of ids/ips technologies," in *Proc. of FIRST*. Citeseer, 2007.

[20] S. Bhattacharya and S. Selvakumar, "Ssenet-2014 dataset: A dataset for detection of multiconnection attacks," in *2014 3rd International Conference on Eco-friendly Computing and Communication Systems*. IEEE, 2014, pp. 121–126.

[21] W. Haider, J. Hu, J. Slay, B. P. Turnbull, and Y. Xie, "Generating realistic intrusion detection system dataset based on fuzzy qualitative modeling," *Journal of Network and Computer Applications*, vol. 87, pp. 185–192, 2017.

[22] "Vmware sdn," *https://lenovopress.com/lp0661.pdf*, July 2020.

[23] "Vmware nfv," *https://www.vmware.com/content/dam/digitalmarketing/vmware/en/pdf/products/nfv/vmware-vcloud-nfv-vcloud-director-edition-datasheet.pdf*, July 2020.

[24] "Public hive mqtt broker," *https://www.hivemq.com/public-mqtt-broker/*, January 2020.

[25] "Public php vulnerable website," *http://testphp.vulnweb.com/*, July 2020.

[26] "Ostinato traffic generator," *https://ostinato.org/*, July 2020.

[27] "Dvwa web service," *http://www.dvwa.co.uk/*, July 2020.

[28] "Owasp security shepherd," *https://owasp.org/www-project-security-shepherd/*, July 2020.

[29] "Metasploitable3 vm," *https://github.com/rapid7/metasploitable3*, July 2020.

[30] "Atop tool," *https://linux.die.net/man/1/atop*, July 2020.

[31] "Nessus tool," *https://www.tenable.com/products/nessus*.

[32] "Nmap tool," *https://nmap.org/*, July 2020.

[33] "Metasploit for eternalblue," *https://null-byte.wonderhowto.com/how-to/exploit-eternalblue-windows-server-with-metasploit-0195413/*, July 2020.

[34] G. M. Kumar and A. Vasudevan, "D-scap: Ddos attack traffic generation using scapy framework," in *Advances in Big Data and Cloud Computing*. Springer, 2019, pp. 207–213.

[35] "Ufonet toolkit," *https://ufonet.03c8.net/*, July 2020.

[36] "The backdoor attack," *https://www.hacking-tutorial.com/hacking-tutorial/5-steps-to-set-up-backdoor-after-successfully-compromising-target-using-backtrack-5/sthash.QgjjbNYM.ULO04i1b.dpbs*, July 2020.

[37] "Hydra tool," *https://tools.kali.org/password-attacks/hydra*, July 2020.

[38] "Cwel tool," *https://tools.kali.org/password-attacks/cewl*, July 2020.

[39] "Man-in-the-middle ettercap," *https://www.1337pwn.com/how-to-perform-a-man-in-the-middle-attack-using-ettercap-in-kali-linux/*, July 2020.

[40] "Cross-site-script toolkit," *https://xsser.03c8.net/*, July 2020.

[41] M. Ahmed, A. Anwar, A. N. Mahmood, Z. Shah, and M. J. Maher, "An investigation of performance analysis of anomaly detection techniques for big data in scada systems," *EAI Endorsed Transactions on Industrial Networks and Intelligent Systems*, vol. 15, no. 3, 5 2015.

[42] M. Ahmed, A. Naser Mahmood, and J. Hu, "A survey of network anomaly detection techniques," *J. Netw. Comput. Appl.*, vol. 60, no. C, pp. 19–31, Jan. 2016.

[43] M. Ahmed, A. N. Mahmood, and M. R. Islam, "A survey of anomaly detection techniques in financial domain," *Future Generation Computer Systems*, vol. 55, pp. 278 – 288, 2016.

[44] Markus Hofmann and Ralf Klinkenberg. *RapidMiner: Data Mining Use Cases and Business Analytics Applications*. Chapman & Hall/CRC, 2013.